\documentstyle[aps,preprint,epsfig]{revtex}
\begin{document}

\title{Nonuniversality and critical point shift in systems with infinitely many absorbing configurations}

\author {Ronald Dickman$^{1,2}$}
\address{
Departamento de F\'\i sica, ICEx,
Universidade Federal de Minas Gerais, Caixa Postal 702,
30161-970, Belo Horizonte - MG, Brasil}
\maketitle
\begin{abstract}
A detailed study of critical spreading in the one-dimensional
pair contact process is performed using a recently devised
reweighting method.  The results confirm the validity of a
generalized hyperscaling relation among 
the (nonuniversal) spreading exponents, and support the assertion that
the critical point location does not depend on the particle density $\phi$.
It appears that the exponents $z$ and $\delta+\eta$, once thought to be
invariant, exhibit a systematic dependence on $\phi$.
\vspace {0.3truecm}

\end{abstract}
\vspace{1em}

\noindent 
$^1${\small electronic address: dickman@fisica.ufmg.br } \\
$^2${\small On leave of absence from: Department of Physics and Astronomy,
Herbert H. Lehman College, City University of New York,
Bronx, NY, 10468-1589.} \\
 
\newpage

\section{Introduction}

The model studied in this work is a not-to-distant cousin of directed percolation.
In view of his extensive and continuing contributions to percolation theory, it gives me
great pleasure to dedicate this work to my friend George Stell, who has taught me 
much about physics (and perhaps more about music).

When viewed as a spatio-temporal process (the oriented axis representing time),
directed percolation exhibits an absorbing-state phase transition, that is,
a transition between an active state and one in which the dynamics is frozen.
Critical phenomena at absorbing-state phase transitions are of longstanding interest in 
statistical physics \cite{privbook,marro},
being found, for example, in models of epidemics \cite{harris},
catalytic kinetics \cite{zgb,evans}, surface growth \cite{alon}, and
self-organized criticality \cite{gz,maslov,dvz,vdmz}.
The best studied models of this kind, 
the contact process (CP) and
directed percolation (DP),
are relevant to experiments on interface pinning \cite{bara}
and sand flow \cite{haye}. 
The transition between active and absorbing states arises out of a conflict between two
opposing processes (e.g., creation and annihilation); when continuous
(as is often the case), it falls 
generically in the DP universality class \cite{janssen,gr1,glb}.
When two or more absorbing states exist and are connected by a symmetry
operation, as in branching and annihilating random walks,
the asymptotic dynamics is dominated by a domain-growth process, and
a new kind of critical behavior appears
\cite{gkt,baw1,iwanbaw,cardybaw}.  
 
Unusual critical behavior also 
appears in models that can become trapped in one of an infinite
number of absorbing configurations (INAC).
(More precisely, the number of absorbing configurations grows exponentially
with the system size. There is no special symmetry 
linking the different absorbing  configurations.)
Models of this sort were introduced to describe surface 
catalysis \cite{benav,albdd,yaldram}; their critical properties have been studied in detail 
by various workers \cite{ijnoco,pcp1,pcp2,mendes,inas,snr,mgd,gcr}.
In one dimension, the pair contact process (PCP) \cite{pcp1}, and other 
models with INAC exhibit static critical behavior in the DP class 
\cite{pcp2,rdjaff}, but the critical exponents $\delta$, $\eta$,
and $z$, associated with the spread
of activity from a localized seed are {\it nonuniversal}, varying 
continuously with the particle density $\phi$ in the 
environment \cite{pcp2,mendes,gcr,lopez}.
[These exponents are defined via the asymptotic 
($t \rightarrow \infty$) power laws:
survival probability $P(t) \sim t^{-\delta}$, mean activity $n(t) \sim t^{\eta}$,
and mean-square distance from the seed $R^2 (t) \sim t^z$.]
This anomalous aspect of critical spreading for INAC can be
traced to a long memory in the dynamics of the order parameter, $\rho$,
arising from a coupling 
to an auxiliary field (the local particle density, $\phi$), that remains frozen in regions where
$\rho = 0$ \cite{inas,mgd,gcr}.  
Theoretical understanding of models with INAC remains incomplete.
Mu\~noz et al. were able to construct a plausible field theory 
(a set of stochastic partial differential equations),
for models with INAC, and to show that the static critical behavior is that of DP \cite{inas}.
Formally eliminating the auxiliary field, they obtained a closed equation for the order parameter,
in which a memory term appears; simulations of this theory
reproduce the nonuniversal exponents observed in
simulations of particle models \cite{lopez}.  On the other hand,
the phenomenon of nonuniversal spreading exponents has so far
resisted analysis via renormalization group or other theoretical
methods.

Studies of spreading in models with INAC have yielded several clues
toward understanding nonuniversality \cite{pcp2,mendes}. First, the {\it location} of the
critical point (the critical parameter value, $p_c$ in the PCP), does not depend on $\phi$, even as the exponents vary.
Second, the exponent $z$, and the sum $\delta + \eta$, are roughly independent
of $\phi$.  Third, the exponents $\delta$, $\eta$, and $z$ obey a generalized form
of the hyperscaling relation derived by Grassberger and de la Torre for models 
with a unique absorbing configuration \cite{mendes,mgt}:
\begin{equation}
\eta + \delta + \frac{\beta}{\nu_{||}} = \frac{dz}{2} \;.
\label{ghypsc}
\end{equation}
For models with a unique absorbing configuration, the scaling relation
$\delta = \beta/\nu_{||} $ holds, and one recovers the original
hyperscaling relation \cite{torre}.  When the static critical behavior
falls in the DP universality class, $\beta/\nu_{||} = \delta_{DP}$ independent
of $\phi$; in what follows we shall write Eq. (\ref{ghypsc}) in the form
$\eta + \delta + \delta_{DP} = dz/2 $.
Recently, \'Odor et al. reported a weak
dependence of $p_c$ on the
particle density in the environment \cite{odor}.
In this work I describe extensive simulations of the PCP, using a
recently-devised reweighting method \cite{rew}, and improved data analysis.
The results shed light on all three of the above-mentioned clues.

The balance of this paper is devoted to
defining the model and simulation algorithm (Sec. II), 
some observations on the implications of a critical point shift for scaling (Sec. III),
outlining the simulation method (Sec. IV), and presenting the simulation
results (Sec. V).  We close (in Sec. VI) with a discussion of our findings
and their relation to previous work on nonuniversal spreading in models
with an infinite number of absorbing configurations.

\section{Model and Scaling Properties}

Jensen's pair contact process (PCP) \cite{pcp1}, is an
{\em interacting particle system}: a Markov process
whose state space is a set of particle configurations
on a lattice \cite{liggett,konno}.  Here we consider the one-dimensional
version: each site of ${\cal Z}^d$ is either vacant or occupied.
Each nearest-neighbor (NN) pair of occupied sites (``particles") has a rate $p$
of mutual annihilation, and a rate $1-p$ of attempted creation.
In a creation event involving particles at sites $i$ and $i+1$,
a new particle may appear (with equal likelihood) either at site $i-1$ or
at $i+2$, provided the chosen site is vacant.
(Attempts to place a new particle at an occupied site fail.)
In an annihilation event the sites occupied by a NN pair are simply vacated.
The PCP exhibits an active phase for $p < p_c$;
above this value the system falls into an
absorbing configuration devoid of NN pairs,
but that typically contains a substantial
density, $\phi$, of isolated particles. The most precise
available estimate for the critical parameter is $p_c = 0.077090(5)$ \cite{rdjaff}.
(Here and in what follows, numbers in parentheses denote an uncertainty
estimate in the last figure or figures.)  Note that the active regime corresponds
to $p < p_c$.

Previous studies leave little doubt that the static critical behavior of the PCP belongs to the
universality class of directed percolation.  Jensen and Dickman found that
the critical exponents $\beta$, $\gamma$, $\theta$ (which govern, respectively,
the stationary mean of, variance of, and initial decay of the order parameter), and
$\nu_{||}$ and $\nu_{\perp}$ (which govern the divergence of the correlation time and
correlation length as one approaches the critical point),
are all consistent with DP values \cite{pcp2}.  More recently, the order parameter moment
ratios and cumulants were found to be the same as those of other models belonging
to the DP universality class \cite{rdjaff}.  
The process of {\it critical spreading}, i.e., the propagation of 
initially localized activity in a system at the critical point, is more subtle.
Critical spreading in the PCP is studied
by simulating the process at the critical point, with an initial condition generated by placing a
single NN pair of particles (the ``seed") into a configuration otherwise devoid of pairs.
In this work, as in previous studies, the sites outside the seed bear a
uniform particle density $\phi$.  (The particles may be placed in a random or a regular fashion.)
The critical point for spreading, $p_c (\phi)$, is defined operationally as the value for
which the survival probability, mean number of pairs, and mean-square distance
from the seed follow power laws.  All earlier studies, except that of \'Odor et al., have found
$p_c (\phi) = p_c$, independent of $\phi$.

The surprising result of the critical spreading studies is that only for a particular value of $\phi$,
the so-called {\it natural density} $\phi_{nat}$, do the spreading exponents $\delta$, $\eta$ and
$z$ assume DP values \cite{pcp2}.  The natural density is defined as the mean particle density
in absorbing configurations generated by the process itself, at $p_c$, starting from a
homogeneous (e.g., fully occupied) configuration.  One may, equivalently, define
$\phi_{nat}$ as the particle density in the critical stationary state, in the thermodynamic limit.
An environment with $\phi > \phi_{nat}$ should favor spreading 
(and one with $\phi < \phi_{nat}$ should hinder it), since the
higher the particle density in the environment, the more pairs will be formed per
creation event.

A kind of spreading phenomenon also arises in the stationary state
due to spontaneous fluctuations.
In the critical stationary state, we can expect to find
inactive regions of all sizes; the particle density 
in large inactive regions is $\phi_{nat}$.  When activity spreads  
into such regions, it should follow the same scaling
behavior as critical spreading with $\phi_{nat}$.  Since the exponents
governing survival and growth of activity in the stationary state 
are subject to the scaling relations $\delta = \beta/\nu_{||}$ and $z = 2\nu_{\perp}/\nu_{||}$,
with $\beta$, $\nu_{\perp}$ and $\nu_{||}$ taking DP values in the stationary state,
it follows that the spreading exponents take their usual DP values as well, for
$\phi = \phi_{nat}$.
In an environment with $\phi \neq \phi_{nat}$, the advance of the active region is
no longer equivalent to the spread of activity in the stationary state, and the spreading
exponents are not constrained to take DP values.
Of course, the interior of the active region
must eventually relax to the critical stationary state, with a particle density
$\phi_{nat}$, regardless of the exterior particle density $\phi$.

In simulations one finds that
for $\phi < \phi_{nat}$ the exponent $\delta$ is larger than the DP value, so that
the survival probability decays more rapidly.  $\eta$ and $z$ are smaller than the
corresponding DP values, showing that diminishing the particle density hampers spreading,
as expected.  For $\phi > \phi_{nat}$ these trends are reversed.  The variations
in $\delta$ and $\eta$ are quite dramatic: the former changes by more than a factor of four,
and the latter by about two, as $\phi$ is varied between the extreme values of 0 and $1/2$.
The changes in $z$ and $\delta+\eta$ are much smaller,
amounting to about 5\% over the full range of $\phi$; it was suggested
that the observed variations, which were not much larger than numerical uncertainties,
could be attributed to corrections to scaling, suggesting the appealing simplification
that $z$ and $\delta +\eta$ are in fact independent of $\phi$ \cite{mendes}.
Since $z$ and $\delta + \eta$ have
only to do with surviving trials (the latter governs the growth in the
number of pairs in such trials: $n_s \sim t^{\delta+\eta} $), lack of
dependence of these quantities on $\phi$ would imply
that the latter only affects the survival probability, and not the asymptotic scaling
of surviving trials.

Two observations should be made before discussing further details.  First,
the above pattern of scaling properties (DP static behavior, nonuniversal exponents
taking DP values only at the natural density, no apparent shift in $p_c$, and near-constancy
of $z$ and $\delta +\eta$), has been
confirmed in other one-dimensional models with INAC (the so-called dimer reaction \cite{pcp2}, the threshold transfer process \cite{mendes}, and an evolution model with real-valued
site variables \cite{lipowski}), in models with a long memory of the initial
configuration \cite{mgd,gcr}, and in a field theory intended to describe
models with INAC \cite{inas,lopez}.  Second, spreading in two-dimensional models with INAC
appears to be more complex \cite{mgd}, and may not exhibit power-law scaling for all 
initial conditions \cite{gcr}, although the static critical behavior again falls in the DP
class \cite{snr,jaffrd}.

\section{Critical Point Shifts and Scaling}

How will shift in the critical point affect the spreading process?
Before attempting an answer, it is helpful define the critical point with somewhat 
greater precision.
In the PCP the {\it bulk} critical point $p_c$ is defined such that 
in the infinite-size limit, the stationary order 
parameter (density of NN pairs) is zero for $p \geq p_c$ (subcritical regime) and scales as
$\rho \sim \Delta  ^{\beta} $ for $\Delta \equiv p_c - p \stackrel {>}{ \sim} 0$.
Now consider a spreading process, in which activity is initially localized (for example,
at a single NN pair in the PCP).  The subcritical regime for spreading is defined as
the set of $p$ values for which the survival probability $P(t) \rightarrow 0$ as 
$t \rightarrow \infty$; the critical value for spreading is then
 
\begin{equation}
p_c(\phi) \equiv \min\{p : \lim_{t \rightarrow \infty} P(t) = 0\} \;.
\label{pcs}
\end{equation}

In simulations, $p_c(\phi)$ is identified as the value for which the survival probability,
mean population $n(t)$, and mean-square spread $R^2 (t)$ all follow 
asymptotic power laws, which define the exponents $\delta$, $\eta$ and $z$.  
(In the PCP, $n(t)$ and $R^2$ refer to NN pairs; in the simpler CP, of course,
they represent the number and spread of particles.
Note that in the supercritical regime, $n \sim t^{d}$ and $R^2 \sim t^2$, that is,
the finite fraction of trials that survive spread with a finite velocity, and have a finite
bulk density.  For $\Delta \stackrel > \sim 0$, we expect a crossover from
the exponents $\eta$ and $z$ at short times to the supercritical growth laws at
longer times.)
We are not aware of any proof that $P$, $n$, and $z$ follow
power laws at $p_c(\phi)$.  But it is known that in the CP, away from the critical 
point, the approach to the stationary state, be it active or absorbing, 
is exponential \cite{liggett,durrett}; in some simpler
cases, such as compact directed percolation, a power-law can be demonstrated explicitly \cite{dk,cdp}.  (On the other hand, the CP with quenched disorder appears to have a 
{\it logarithmic} time-dependence at the critical point, and power-law away from the critical
point \cite{dcp1,dcp2}).  In any event, we shall assume that in the PCP, $p_c(\phi)$ 
can be identified via the power-law criterion.

Consider the mean population $n_s (t)$, 
in trials that survive until (at least) time $t$.  By definition, $n_s (t) \geq 1$, for
{\it any} value of $p$, but for  $p > p_c(\phi)$, $n_s (t)$ remains bounded as
$t \rightarrow \infty$, while for $p < p_c(\phi)$ it grows without bound.  Unbounded
growth also obtains at $p_c(\phi)$ if we assume
that $P(t)$ decays slower than exponentially at the critical point
(absenting a proof, we merely observe that exponential decay would be
incompatible with scale-invariance!).  
That is because
the rate of extinction is proportional to the probability of having exactly one
pair:

\begin{equation}
- \frac{dP}{dt} = p P(t) \Pr \;[n_s (t) \!=\! 1]  \;.
\label{extinct}
\end{equation}
If $n_s (t)$ remains bounded, then $\Pr \;[n_s (t) \!=\! 1] > 0$, and Eq. \ref{extinct}
implies exponential decay of the survival probability.  Assuming the power laws
noted above for $P$ and $n$, we have $n_s (t) \sim t^{\eta + \delta}$;
unbounded growth of $n_s (t)$ at $p_c (\phi)$ implies $\eta + \delta \geq 0$.

The above observations are helpful in analyzing the implications of a critical point shift.
Suppose that $p_c(\phi) < p_c $ for spreading in an environment with
$\phi < \phi_{nat}$.  (That is, to compensate for the ``hostile" environment, the spreading critical
point shifts to a value that lies in the bulk {\it supercritical} regime.)
This means that for $ p_c(\phi) < p < p_c$, the bulk is
supercritical, but the spreading process always dies.  
At the critical point $p_c(\phi)$, the spreading trials
that do survive for long times have a nonzero bulk density
$\rho_b \sim [p_c - p_c(\phi)] ^{\beta}$, since the interior of the active region must eventually
attain the stationary state.  At long times, this interior region, having a finite density, makes
the dominant contribution to $n_s$ and $R^2$, implying that 
$n_s \sim (R^2)^{d/2} \sim t^{zd/2}$, which in turn implies the hyperscaling relation
\begin{equation}
\eta + \delta = \frac{dz}{2} \;,
\label{scalefd}
\end{equation}
rather than Eq. (\ref{ghypsc}).  [Eq. (\ref{scalefd}) describes spreading at first-order
transitions between different absorbing states, for example in compact DP or the voter model \cite{hypdk}.  In this case Eq. (\ref{scalefd}) follows directly from Eq. (\ref{ghypsc}), since $\beta = 0$.]

Suppose, conversely, that $p_c(\phi) > p_c$ for $\phi > \phi_{nat}$, i.e., the ``friendly"
environment permits a spreading trial to survive indefinitely even if the bulk is subcritical.
If so, then for $p_c < p \leq p_c(\phi)$, a surviving trial must have its activity
concentrated in an annular region (or at two fronts, in one dimension), since the
order parameter vanishes in the bulk.  (One may think of the activity as a
``chemical wave" that converts one kind of absorbing configuration, with $\phi > \phi_{nat}$, 
into another, with $\phi \simeq \phi_{nat}$.)
In one dimension, moreover, the width of this
active zone must grow fast enough that $n_s (t)$ grows without bound as
$t \rightarrow \infty$.

Summarizing, if the critical point for spreading depends on the surrounding
particle density, 
then we should expect, for $\phi < \phi_{nat}$, that
the activity profile $\rho_s (x,t)$ in surviving trials is plateaulike, and that the
spreading exponents satisfy Eq. (\ref{scalefd}); for $\phi > \phi_{nat}$, we would
expect the activity profile to be bimodal, with an inactive central region.  There are no arguments, to our knowledge, prohibiting such behaviors.  (Indeed,
such modifications of usual spreading were reported 
for two-dimensional models in Refs. \cite{snr} and \cite{mgd}.)  
But if the activity profiles do not take the anticipated forms, it is likely that either no
shift in $p_c(\phi)$ exists, or that the asymptotic long-time behavior has not been probed.

\section{Simulation Method}

{\it Initial configuration.}  $\:$ For each trial, we construct
an initial configuration with a single NN pair or ``seed" at the center of the system
(sites $L/2$ and $L/2 +1$ on a line of $L$ sites).  We consider three
values of $\phi$: zero, 1/2, and the natural density 0.242.
For $\phi = 1/2$ the sites immediately to either side of the seed are vacant,
beginning an alternating sequence that frames seed.
For the studies at $\phi_{nat} $ we generate random initial configurations as follows.
Consider the sites $i \geq L/2 + 1$.  Since $L/2 +1$ is occupied, it must be followed
by a gap of $g \geq 1$ vacant sites before the next occupied site.  We generate a
sequence of gaps $g_i = 1 + m_i$ where the $m_i$ are a set of independent Poisson
random variables with mean $\langle m_i \rangle = \phi_{nat}^{-1} - 2$,
in order to reproduce the natural particle density.
(The same procedure is applied to sites $L, L-1,...1$.
In each trial we keep a record of the extent of the active region.  To prepare for
the following trial we only need to reset sites within that region; since it is usually
much smaller than the full system, this results in a considerable speedup.)

{\it Dynamics.} $\;$ We maintain a list of the $N_p$ current nearest-neighbor
pairs.  At each step we choose a pair at random from the list, and a
process (annihilation with probability $p$, creation with probability $1-p$).
In case of annihilation, the pair is simply removed.  For creation, 
we choose a site $i$ at random from the two sites neighboring of the pair, 
and place a new particle at $i$ if it is currently vacant.  
(If $i$ is occupied the configuration remains the same.)  
The time increment associated with a step is $\Delta t = 1/N_p$,
corresponding to one transition per pair per unit time, in agreement with 
the transition rates that define the process.
Following each change we update the list of pairs.
The trial continues until either no pairs remain (extinction) or a preset
maximum time $t_m$ is reached.  (In the largest studies reported here,
$t_m = 22026 \simeq e^{10}$.)
We use a  lattice of $L = 5 \times 10^4$ sites, sufficiently large that activity
never reaches the boundary of the system for $t \leq t_m$. 

{\it Reweighting scheme.} We have realized a 
significant speedup, permitting
larger samples and longer durations than in previous studies, by means of
a reweighting technique.  The basis and details of this method are given in
Ref. \cite{rew}.  Briefly, we run the simulations at a central value, $p_0$, of
the annihilation probability $p$, and reweigh the sample to study nearby values
$p' = p_0 + n \Delta p$.  (We always use $p_0$ very near $p_c$, typically
$p_0 = 0.07709$ or 0.0771; $\Delta p$ is either $10^{-5}$ or $5 \times 10^{-6}$,
depending on $t_m$.)  Suppose that in a particular trial,
there have been $n_a$ annihilation events and $n_c$ creation events (successful or
not) up to time $t$.  The weight for this sequence of events in a
simulation with annihilation probability $p'$, is

\begin{equation}
\omega (p',p_0;t) = \left(\frac {p'}{p_0} \right)^{n_a}  \left(\frac {1-p'}{1-p_0} \right)^{n_c} \;,
\label{omega}
\end{equation}

\noindent which represents the ratio of the probability for this sequence in the process with
parameter $p'$ to the corresponding probability for $p_0$.
In practice, we record the number of pairs, mean-square displacement, and 
continued survival of a trial at unit time intervals.  The simulation estimate for
the central value, $p_0$, of any property $A$ at time $t$ is, as usual, 
$\overline{A}_{t;p_0} \equiv N_{tr}^{-1} \sum_{k=1}^{N_{tr}} A_{k,t}$,
where $A_{k,t}$ is the value at time $t$ in the $k$th of a total of $N_{tr}$ independent trials.
The corresponding estimate for parameter value $p'$ is
$\overline{A}_{t;p'} \equiv N_{tr}^{-1} \sum_{k=1}^{N_{tr}} \omega (p',p_0;t) A_{k,t}$.
The reweighting factor $\omega$ varies from trial to trial, depending on the number of
creation and annihilation events realized up to time $t$.  An analysis of the
range of $p$ values for which reweighting is useful is given in Ref. \cite{rew}.

Application of this method to
the CP yielded results of unprecedented precision for critical point and exponents \cite{rew}.  
Beyond the obvious economy of studying various parameter values at once,
the fact that the results for all $p$ values come from the same data set
eliminates run-to-run fluctuations that tend to obscure the value of $p_c$.
(Our method, of course, does not eliminate fluctuations {\it per se}.  But here the
fluctuations affect the results for all parameter values uniformly.  Without reweighting,
one has to generate a different sample, with its particular fluctuations, for each
parameter value of interest.)

\section{Results}

\subsection{Critical Point and Spreading Exponents}

We determine the survival probability, mean number of pairs, and
mean-square distance of pairs from the original seed, expecting that these follow
asymptotic power laws at the critical point, and show deviations from power laws
for off-critical values of $p$.
To identify $p_c (\phi)$ we follow the widely-used practice of studying the local slopes, $\delta(t)$, $\eta(t)$ and $z(t)$,
defined as the derivatives of $\ln P$, $\ln n$, and $\ln R^2$, respectively, with
respect to $\ln t$.
We evaluate $\delta(t)$ by performing a least-squares linear fit
to the $\ln P$ data for a set of  25 equally-spaced values (an increment of
0.1) of $\ln t$; $\eta(t)$ and $z(t)$ are obtained similarly.

In studies of spreading at an absorbing-state phase transition, 
the local slopes are usually plotted versus $t^{-1}$.  The curves for $\delta (t)$ and
$\eta (t)$ typically fall into three groups: those that curve sharply upward are
taken as marking supercritical parameter values, those curving sharply downward
are associated with the subcritical regime, and those that seem to have a finite
limiting value (as $t^{-1} \rightarrow 0$) are consistent with criticality.
(The extrapolated value furnishes an estimate for
the associated exponent, $\delta$, $\eta$, or $z$.
There is often some uncertainty, i.e., two or more parameter values may ``look"
critical, especially when fluctuations are strong.  The curves for $z(t)$ generally
provide much less information about the location of the critical point.)
The rationale for this
practice is that in the one-dimensional CP the leading correction to scaling
appears to be $\propto t^{-1}$, that is, for $t \gg 1$,

\begin{equation}
\delta(t) \simeq \delta [ 1 + a t^{-\Delta_1} + b t^{-\Delta_2} + \cdots],
\label{corrtosc}
\end{equation}
with $\Delta_1 = 1$.  In other cases, however, the dominant
correction to scaling may decay more slowly than $1/t$ \cite{grass89}.

In Fig. 1 we plot the local slopes $\delta(t)$ from our spreading studies with $\phi = 0$.
(The data are from a set of $2 \times 10^6$ trials, extending to $t_m = 22026$.  The
simulations were performed at $p_0 = 0.07709$, and reweighted to study a series of
$p$ values at intervals of $\Delta p = 5 \times 10^{-6}$.)  When plotted versus $t^{-1}$, 
as in Fig. 1, all of the curves swerve sharply downward; there is no obvious candidate for 
$p_c$.  At this point we recall an important observation of \'Odor
et al.: the particle density $\phi(t)$ in the active region approaches its natural
value as a power law, $|\phi(t) - \phi_{nat}| \sim t^{-\delta_{DP}}$, with the standard
DP exponent $\delta_{DP} \simeq 0.16$ \cite{odor}.  Given the coupling between the order parameter and the particle density, it is natural to suppose that the slow relaxation
of the latter will yield a dominant correction to scaling $\sim t^{-\delta_{DP}}$ rather
that $\sim t^{-1}$.  We test this hypothesis in Fig. 2, where the $\delta(t)$ data are
replotted versus $t^{-\delta_{DP}}$.  Evidently, the strong curvature of the $\delta(t)$
plot in Fig. 1 is accounted for by a change in the correction to scaling exponent; the
graphs of Fig. 2 are roughly linear.  Similar changes attend the 
switch from $t^{-1}$ to $t^{-\delta_{DP}}$ in the other local-slope plots, for all of the
$\phi$ values studied here.  

In Fig. 2, we see that the curve for $p = 0.07707$ is nearly linear, whilst those for
0.077065 and 0.077075 show noticeable curvature.  (This visual impression is confirmed
by least-squares quadratic fits to the data.)  Linear extrapolations of the $\delta(t) $
plots yield $\delta = 0.252$, 0.256, and 0.261 for $p= 0.077065$, 0.07707, and 0.077075,
respectively.  Based on these results, we estimate $\delta_{\phi = 0} = 0.256(4)$.
Fig. 3 shows a similar plot for $\eta(t)$, the local slope of the population size.  The curves
are somewhat less regular, and the most linear plot appears to be that for $p= 0.077075$.
Linear extrapolations for $p = 0.07707$, 0.077075, and 0.07708 yield $\eta = 0.218$,
0.211, and 0.204, leading to the estimate $\eta_{\phi = 0} = 0.211(7)$.  The data
for $z(t)$ are plotted in Fig. 4.  Here it is difficult to discern $p_c$;
linear extrapolations over the range $p =0.077065$ -
0.07708 yield $z_{\phi=0} = 1.245(7)$.  Our results are consistent with those
reported in Ref. \cite{pcp2} for spreading in an initially empty lattice:
$\delta_{\phi = 0} = 0.250(5)$, $\eta_{\phi=0} = 0.215(5)$, and $z_{\phi=0} = 1.238(4)$.
(The somewhat larger error bars in the present results reflect the uncertainty
inherent in extrapolating to $t^{-0.16} = 0$.)  Combining the critical point estimates
from the analyses of $\delta(t) $ and $\eta(t)$, we obtain $p_c (0) = 0.077073(6) $.

Next we consider spreading in a half-filled environment.  Here we performed a total of
$1.2 \times 10^6$ trials, extending to $t_m = 22026$; in this case we
used $p_0 = 0.0771$ as the central simulation value.  The local slopes $\delta(t)$
and $\eta (t)$ are shown in Figs. 5 and 6, respectively.  
Note that $-\delta$ and $\eta$ approach their limiting values from below,
while for $\phi = 0$ the approach is from above.
Analyzing the local
slopes as above, we find $p_c (1/2) = 0.077087(6)$.  
The exponent estimates are listed in Table I.  
Finally, we performed simulations of spreading into an environment
with particle density $\phi_{nat} = 0.242$.  The central value of $p$ was again 0.07710,
and the number of trials $2 \times 10^6$, but the studies extended only to time 8103.
Analyses of $\delta(t)$ and $\eta (t)$ (the latter is shown in Fig. 7) 
yield $p_c (\phi_{nat}) = 0.077093(7)$.
The critical exponents for $\phi = \phi_{nat}$, listed in Table I, are in good 
agreement with the known DP values.

\subsection{Activity and particle density profiles}

The discussion of Sec. III motivates our study of the spatial distribution of activity
(the pair density) and the particle density in surviving trials.  According to the 
scaling hypothesis \cite{torre}, at the critical point, the mean order-parameter density $\rho_s(x,t)$ in surviving trials,
given an initial seed at $x=0$ and $t=0$, should scale as

\begin{equation}
\rho_s (x,t) = t^{z/2 - \delta - \eta} \; R ( x^2/ t^z ),
\label{rhosc}
\end{equation}
where the scaling function $R$ vanishes for large values of its argument.
The hyperscaling relation Eq. (\ref{ghypsc}) implies that the prefactor is
$t^{-\delta_{DP}}$.  We therefore expect plots
of $\rho ^* (x^*, t) \equiv t^{\delta_{DP}} \rho (x/t^{z/2},t)$ to collapse onto a single curve.
Figures 8, 9, and 10 show such plots for $\phi = 0$, 1/2, and $\phi_{nat} = 0.242$,
respectively, confirming the anticipated scaling.  
(The study for $\phi = 0$ was performed using $p$ = 0.07708, and those
for $\phi = 0.242$ and 1/2 using $p=0.07709$; there is no rewighting.)
The scaling function has the same qualitative form in all three cases; there is no hint of the
plateaulike or bimodal profiles envisaged in Sec. III.  The fact that the scaled activity
profile is stable for times $\geq 1000$ suggests that the local slopes
$\delta (t)$, $\eta (t)$ and $z(t)$ have also attained their asymptotic scaling regime,
so that the critical exponents can be extrapolated with confidence.

We note also that the rate of
spreading is greater, the larger is $\phi$.  This can be quantified by measuring the
mean-square spread; we find $\overline{x^{*2}} \simeq 0.75 t$, $1.1t$, and $2.0 t$
for $\phi = 0$, 0.242, and 0.5, respectively.  Most significantly, the definition
$x^* \equiv x / t^{z/2}$ involves the $\phi$-{\it dependent} $z$ values listed in Table I.
Were we to use, for example, the standard DP value $z = 1.265$ in 
$x^*$ for $\phi = 0$ we would find that $\overline{x^{*2}} /t$ approaches zero
as $t \rightarrow \infty$.  

A similar analysis can be applied to the mean particle density $\phi_s (x,t)$
in surviving trials.  Here $\phi_s (x,t)$ is fixed at its initial value $\phi$ 
for $x \gg t^{z/2}$, that is, outside the active region, and we expect $\phi_s (x,t) - \phi_{nat}$
to decay $\sim t^{-\delta_{DP}} $ within the active region.
These behaviors are reflected in the scaling form

\begin{equation}
\phi_s (x,t) - \phi_{nat} = (\phi - \phi_{nat}) \left[ 1 - \frac{\rho_s(x,t)}{\rho_s (0,t)} \right]
                                    + t^{-\delta_{DP}} F( x^2/t^z) \;.
\label{psc}
\end{equation}
Here the first term on the r.h.s. represents the uniform particle density $\phi$
outside the active region; the factor in square brackets switches from 1 outside,
to zero well inside this region.  (For $\phi = \phi_{nat}$, of course, this term vanishes.)
The second term represents relaxation of the particle density to its natural value
within the active region.  To test this scaling hypothesis we plot

\begin{equation}
\phi^* (x^*,t) \equiv t^{\delta_{DP}} \left\{ (\phi_s(x,t) - \phi_{nat})
                  -  (\phi - \phi_{nat}) \left[ 1 - \frac{\rho_s(x,t)}{\rho_s (0,t)} \right] \right\} \;,
\label{phistar}
\end{equation}
versus $x^*$.  Figures 11, 12, and 13 are scaling plots of $\phi^* (x^*,t)$ for
$\phi = 0$, 1/2, and 0.242, respectively.  The data collapse confirms the observation
of \'Odor et al. that the particle density exhibits a power-law relaxation to its
natural value, $|\phi (t) - \phi_{nat} \sim t^{-\delta_{DP}}|$, within the active region \cite{odor}.

\section{Discussion}

We return to the three clues mentioned in the Introduction.
First our simulations, which are the most extensive
(in terms of sample size and trial duration),
of the PCP to date, shows no evidence for a shift of the critical
point with particle density.  It is true (see Table I) that our $p_c$ estimate for $\phi = 1/2$
is slightly higher than that for $\phi = 0$, but the data for the three $\phi$ values studied
show no systematic trend.  In particular, our data do not confirm the results reported 
in Ref. \cite{odor}, where $p_c$ is given as 0.07704 for $\phi=0$ and 0.07714 for
$\phi = 0.432$ (and would presumably be at least as large, for $\phi = 1/2$), so that
$p_c(1/2) - p_c(0) \geq 10^{-4}$.
Here, by contrast, we
find $p_c(1/2) - p_c(0) = 1.4 \times 10^{-5}$, 
with an uncertainty of $\pm 1.2 \times 10^{-5}$.  To summarize, the present results
are incompatible with the relatively large shift reported in Ref. \cite{odor},
but are quite compatible with there being no shift at all.

In Figures 14, 15 and 16 we plot our results for the spreading exponents
$\delta$, $\eta$, and $z$ as functions of $\phi$, together with results from
Refs. \cite{pcp2} and \cite{odor}.  The overall consistency among these
studies leaves little doubt that there is a systematic, roughly linear dependence
of the exponents on the particle density.  

Our results for the spreading exponents are sufficiently precise to rule out the
proposal that $z$ and $\delta+\eta$ are independent of $\phi$.  While their
variation is much smaller than that in $\delta$ or $\eta$ separately, 
there are significant changes,
amounting to a relative variation of about 5\% between the extremes $\phi = 0$ and
$\phi = 1/2$.  The fact that $z$ varies with $\phi$ was confirmed in our scaling analysis
of the activity profile.
Finally, as shown in Table I, the generalized hyperscaling relation is satisfied to within
numerical uncertainty.  Given the variation in $z$, a compensating variation in
$\delta + \eta$ is needed for hyperscaling to hold.
That $\delta+\eta$ and $z$ depend on $\phi$ forces us to abandon what seemed
an attractive simplifying hypothesis: the environment affects not only the likelihood
of survival, but (to a lesser extent) the scaling properties of trials that do survive.

Can we understand nonuniversal spreading in models with INAC?  Thus far, attempts to identify
a marginal parameter in a renormalization group analysis of the PCP and allied
models have not borne fruit.  Nonuniversality does appear to be associated
with a long memory, due to the slow relaxation of an auxiliary field coupled
to the order parameter.  Such a memory term appears in a field theory that
successfully predicts the static critical behavior of the PCP \cite{inas}, and
that reproduces (numerically) nonuniversal spreading \cite{lopez}.  Moreover,
varying exponents have been observed in one-dimensional
models with a unique absorbing configuration and a slowly-relaxing memory of
the initial condition \cite{mgd,gcr}.  Explaining the 
nonuniversality of the spreading exponents, and predicting their values as
a function of the particle density, remain as theoretical challenges.

{\bf Acknowledgements}

I am grateful to Alessandro Vespignani and Geza \'Odor for helpful comments.

This work was supported by CNPq and FAPEMIG.

\newpage

\begin{table}
\caption{Critical point $p_c (\phi)$ and spreading exponents for the PCP. $\;$
DP exponents from Ref. [48].
Numbers in parentheses denote uncertainties in the last figure.}
\begin{center}
\begin{tabular}{|r|l|l|l|l|l|l|} 
$\phi$ &$p_c (\phi)$   &  $\delta$ & $ \eta $ &  $z$&$\delta +\eta$ &$2\eta +2(\delta +\delta_{DP}) - z$ \\
\hline\hline
0       & 0.077073(6)  & 0.256(4) &  0.211(7) & 1.245(7) & 0.467(11)  & 0.01(3)   \\
0.242 & 0.077093(7)  & 0.161(2) &  0.314(4) & 1.264(4) & 0.475(6)   & 0.005(16)     \\
0.5    &  0.077087(6) & 0.056(3) &  0.443(5) &  1.303(3) & 0.499(8)   & 0.014(19)    \\
\hline
DP    &                    & 0.1595   &  0.3137   &  1.2652   & 0.4732(1) &                    \\
\hline
\end{tabular}
\end{center}
\label{pcprat}
\end{table}

\newpage

\noindent {\bf Figure Captions}
\vspace{1em}

\noindent Fig. 1. Local slope $-\delta(t)$ versus $t^{-1}$ for the PCP with
$\phi = 0$.  The central curve (with data points) is the simulation result
at $p = 0.07709$; curves above and below were obtained via reweighting,
for intervals $\Delta p = 5 \times 10^{-6}$.  The inset is a detail of the
late-time behavior.
\vspace{1em}

\noindent Fig. 2. The data of Fig. 1 plotted versus $t^{-0.16}$.
\vspace{1em}

\noindent Fig. 3. Local slope $\eta(t)$ for $\phi = 0$, versus $t^{-0.16}$,
curves as in Fig. 1.
\vspace{1em}

\noindent Fig. 4. Local slope $z(t)$ for $\phi = 0$, versus $t^{-0.16}$,
curves as in Fig. 1.
\vspace{1em}

\noindent Fig. 5. Local slope $-\delta(t)$ versus $t^{-0.16}$, for $\phi = 1/2$.
The central curve (with data points) is the simulation result
at $p = 0.07710$; curves above and below were obtained via reweighting,
for intervals $\Delta p = 5 \times 10^{-6}$. 
\vspace{1em}

\noindent Fig. 6. Local slope $\eta(t)$ versus $t^{-0.16}$, for $\phi = 1/2$;
curves as in Fig. 5.
\vspace{1em}

\noindent Fig. 7. Local slope $\eta(t)$ versus $t^{-0.16}$, for $\phi = 0.242$.
The central curve (with data points) is the simulation result
at $p = 0.07710$; curves above and below were obtained via reweighting,
for intervals $\Delta p = 5 \times 10^{-6}$. 
\vspace{1em}

\noindent Fig. 8. Scaled activity density $\rho^* = t^{\delta_{DP}} \rho$ versus
$x^* = x/t^{z/2}$ in the critical PCP with $\phi = 0$.  Open squares: $t = 1000$;
diamonds: $t= 5 000$; grey squares: $t = 10 000$; line: $t = 20 000$.
\vspace{1em}

\noindent Fig. 9. Scaled activity density as in Fig. 8, but for $\phi = 1/2$.  
\vspace{1em}

\noindent Fig. 10. Scaled activity density as in Fig. 8, but for $\phi = 0.242$.  
\vspace{1em}

\noindent Fig. 11. Scaled particle density $\phi^* $ versus
$x^* $ in the critical PCP with $\phi = 0$.  Symbols as in Fig. 8
\vspace{1em}

\noindent Fig. 12. Scaled particle density as in Fig. 11, for $\phi = 1/2$.
\vspace{1em}

\noindent Fig. 13. Scaled particle density as in Fig. 11, for $\phi = 0.242$.
\vspace{1em}

\noindent Fig. 14. Critical exponent $\delta$ versus particle density $\phi$.
Filled squares: present study; open squares: Ref. \cite{odor}; circles: Ref. \cite{pcp2}.
\vspace{1em}

\noindent Fig. 15. Critical exponent $\eta$ versus particle density $\phi$.
Symbols as in Fig. 14.
\vspace{1em}

\noindent Fig. 16. Critical exponent $z$ versus particle density $\phi$.
Symbols as in Fig. 14.
\vspace{1em}

\newpage

\begin{figure} 
\centerline{\epsfig{file=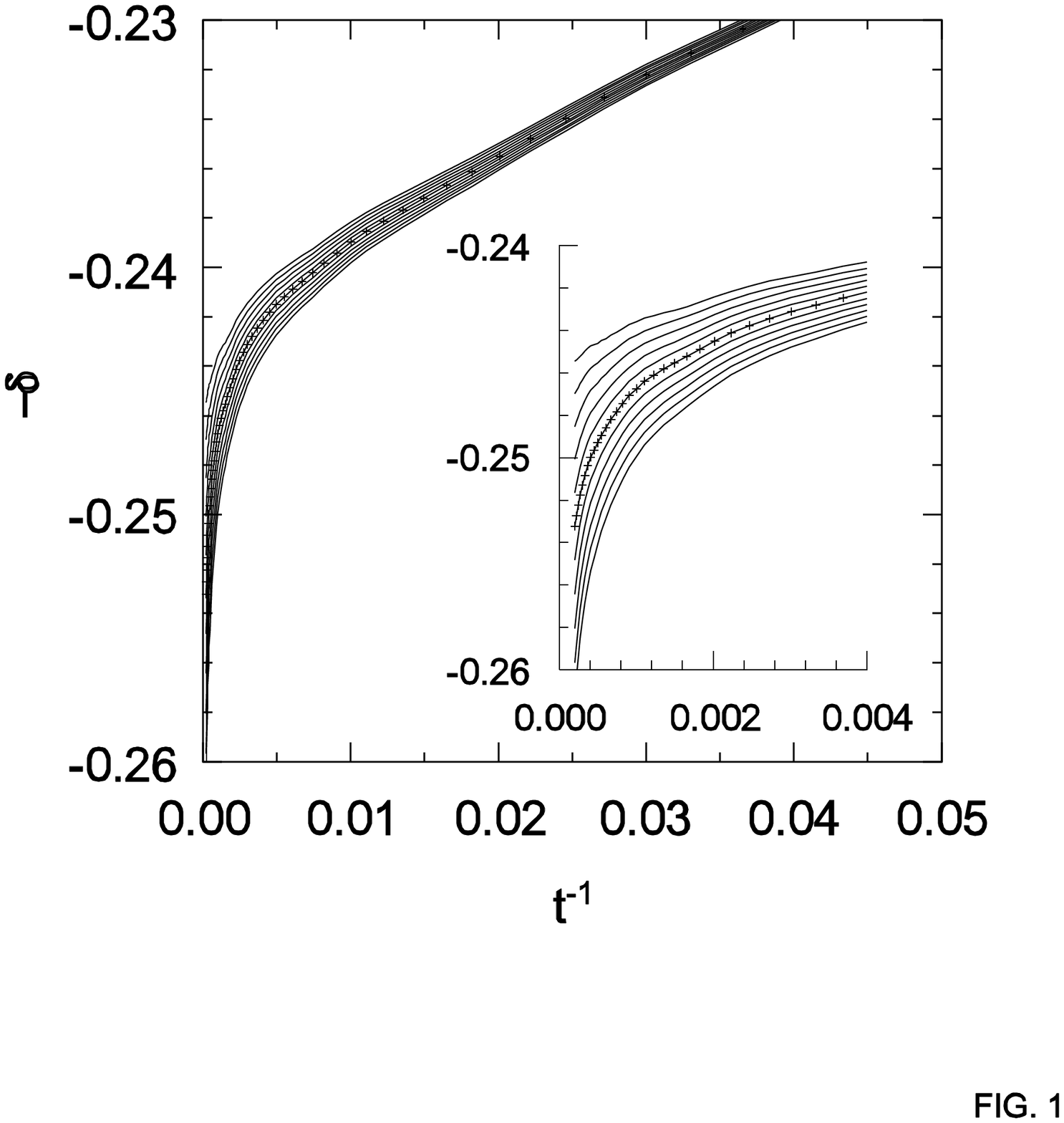,height=6.0in,width=4.5in}}
\end{figure}

\begin{figure} 
\centerline{\epsfig{file=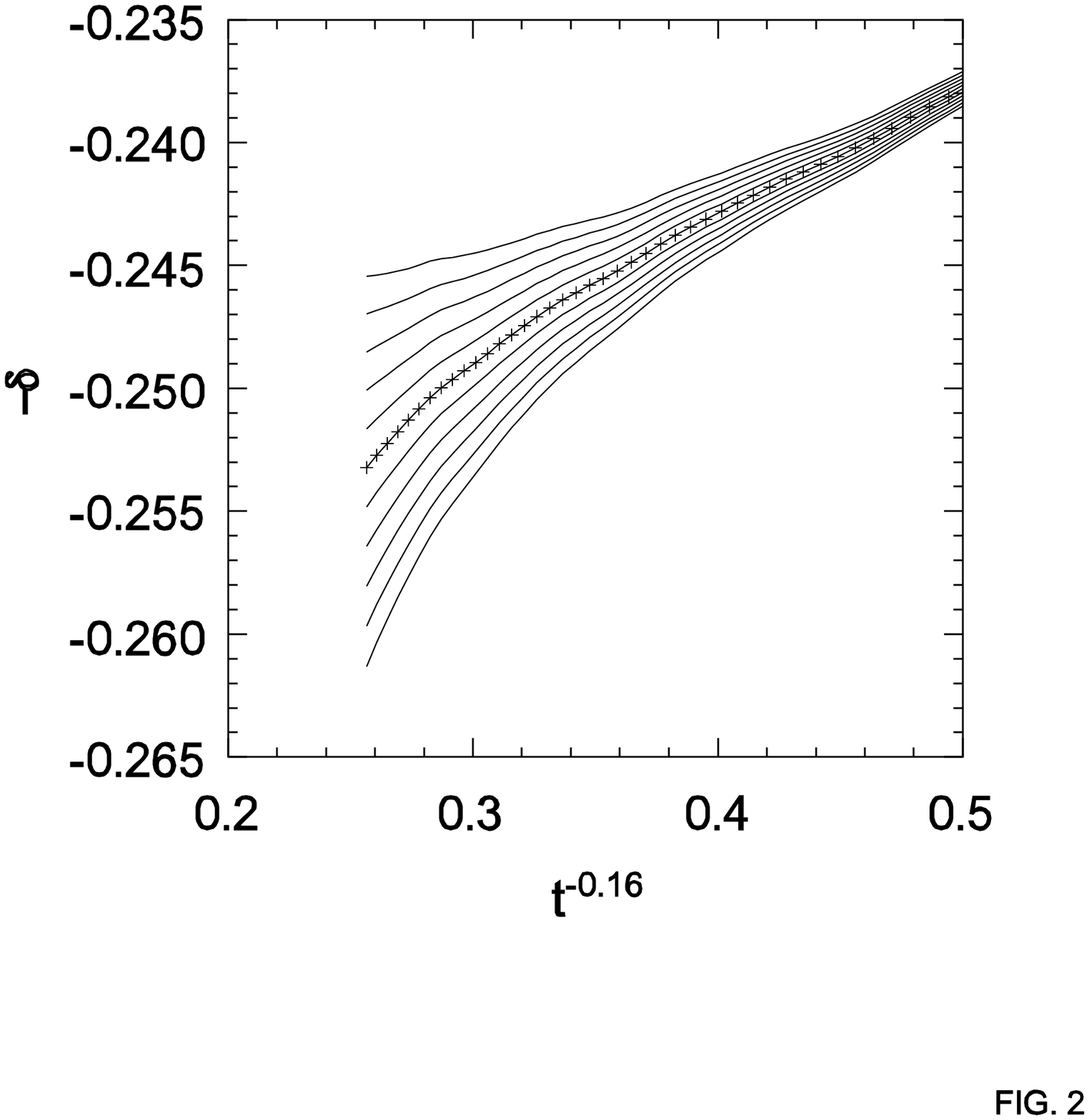,height=6.0in,width=4.5in}}
\end{figure}

\begin{figure} 
\centerline{\epsfig{file=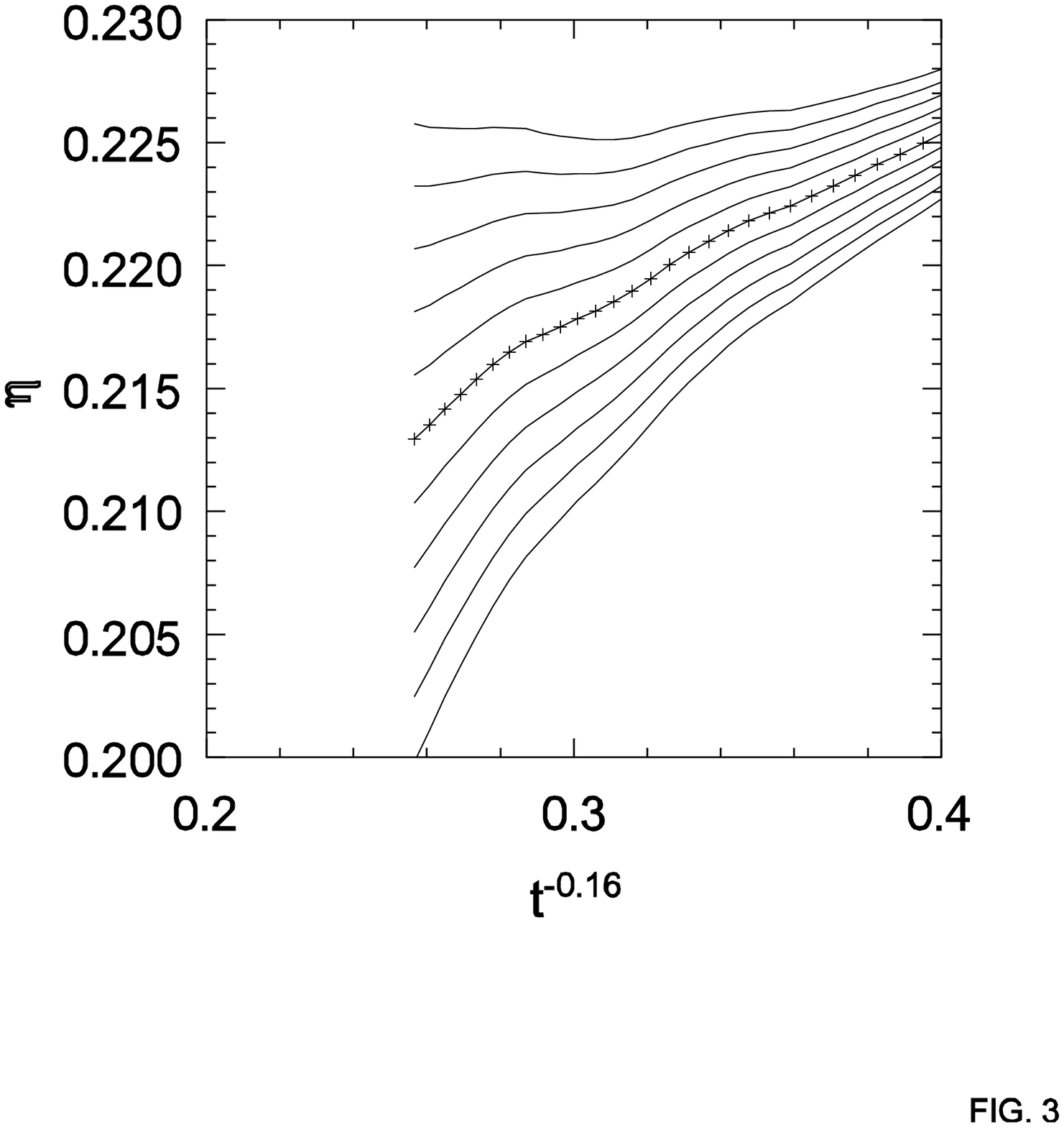,height=6.0in,width=4.5in}}
\end{figure}

\begin{figure} 
\centerline{\epsfig{file=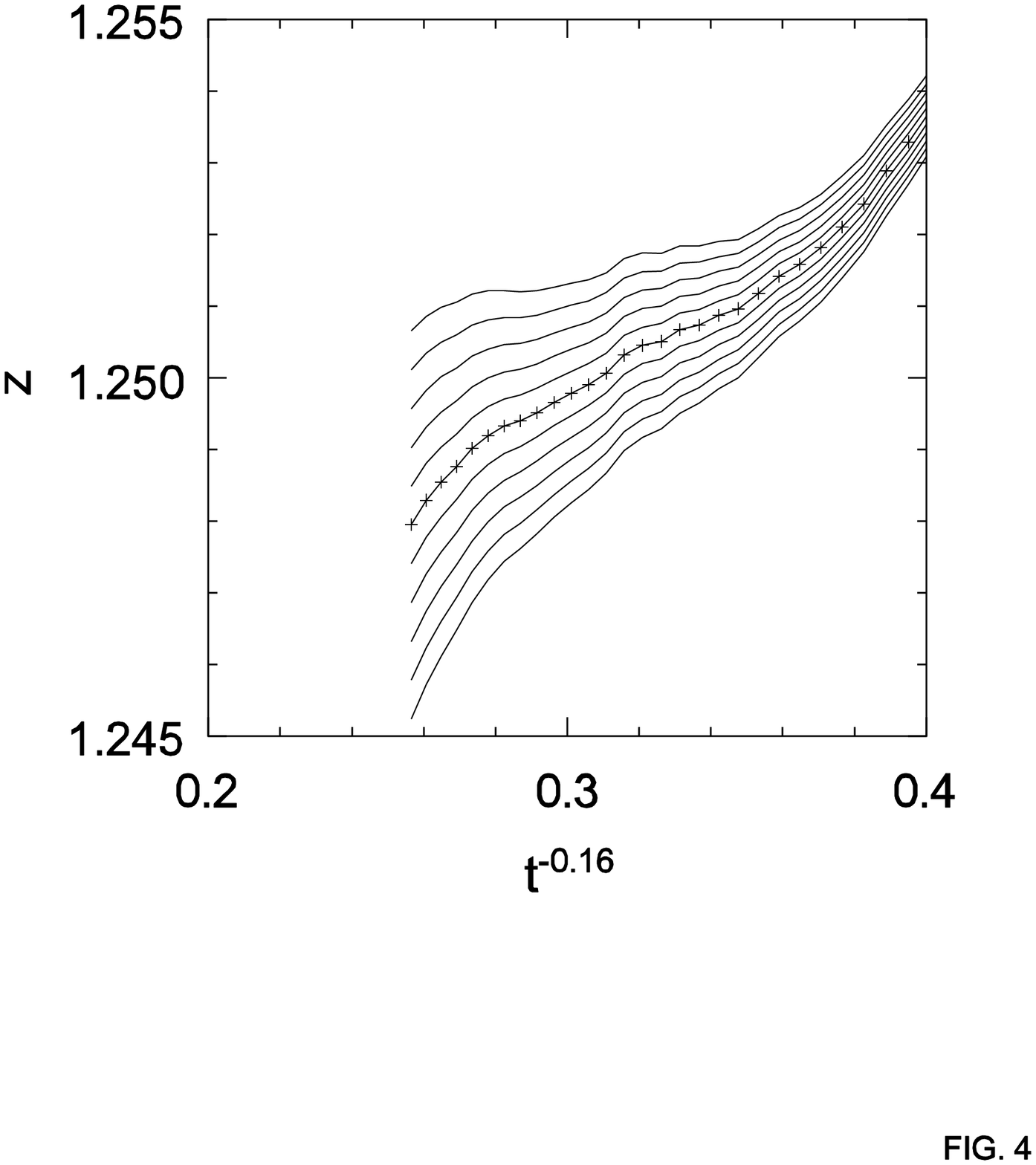,height=6.0in,width=4.5in}}
\end{figure}

\begin{figure} 
\centerline{\epsfig{file=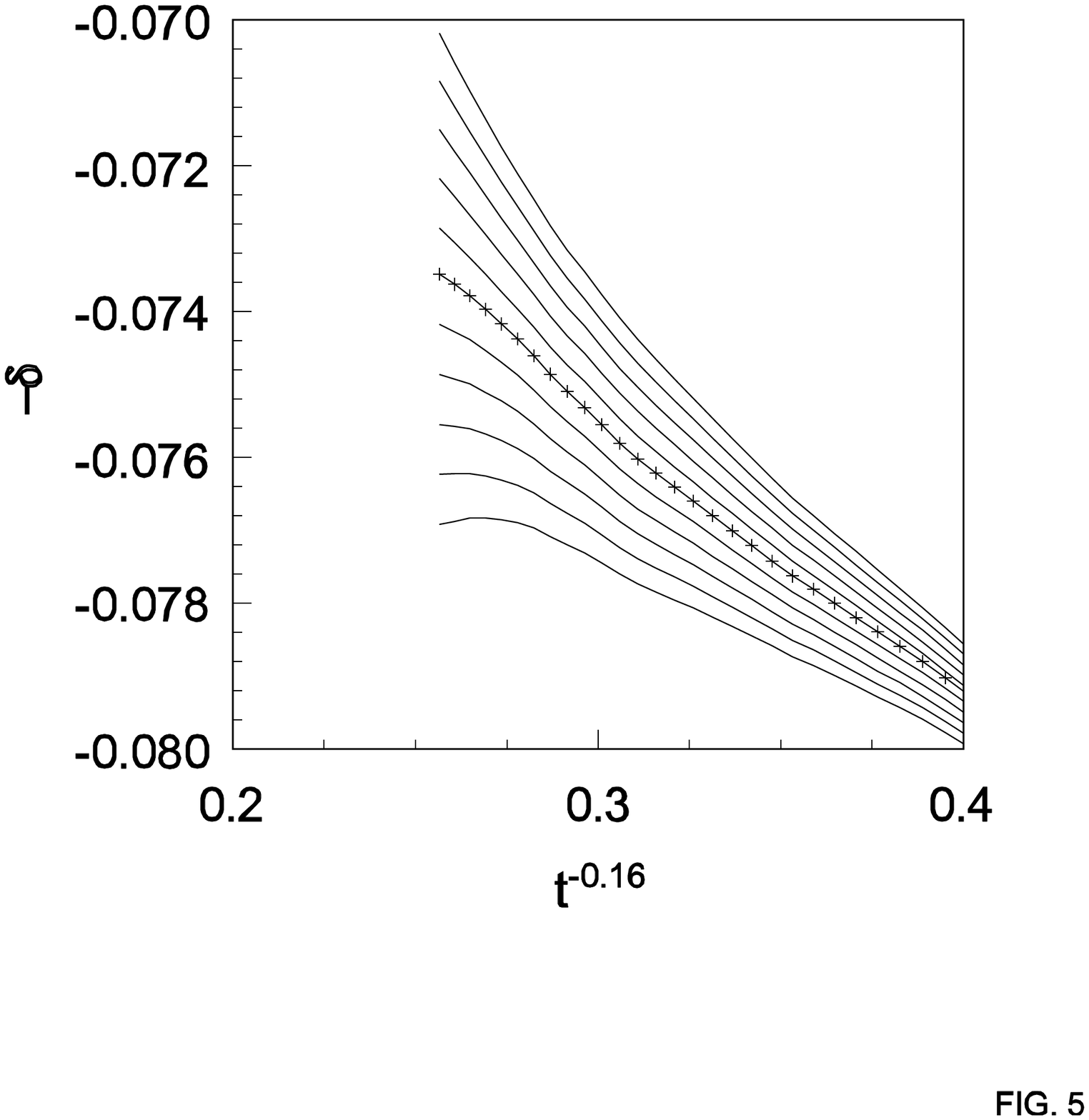,height=6.0in,width=4.5in}}
\end{figure}

\begin{figure} 
\centerline{\epsfig{file=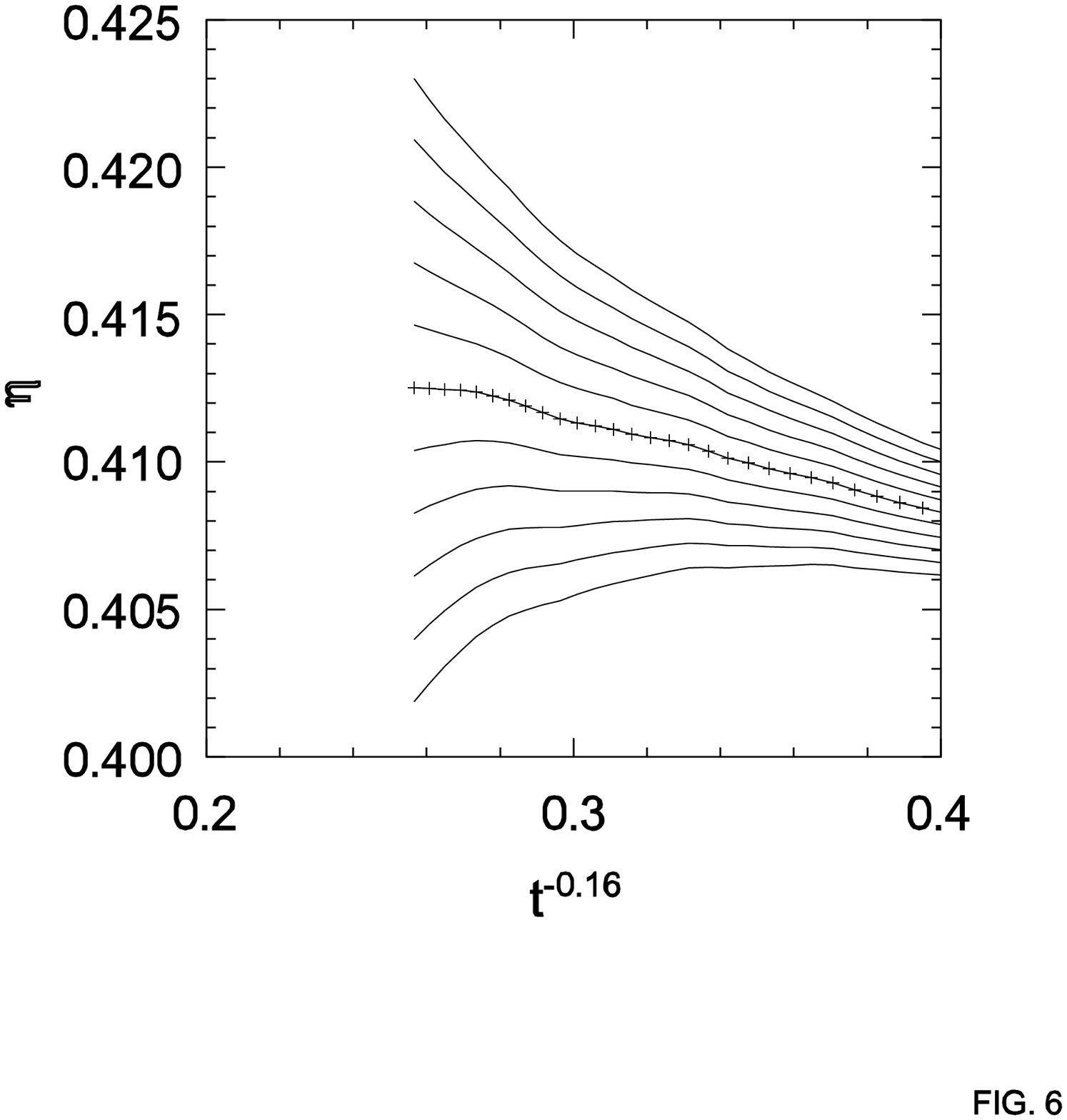,height=6.0in,width=4.5in}}
\end{figure}

\begin{figure} 
\centerline{\epsfig{file=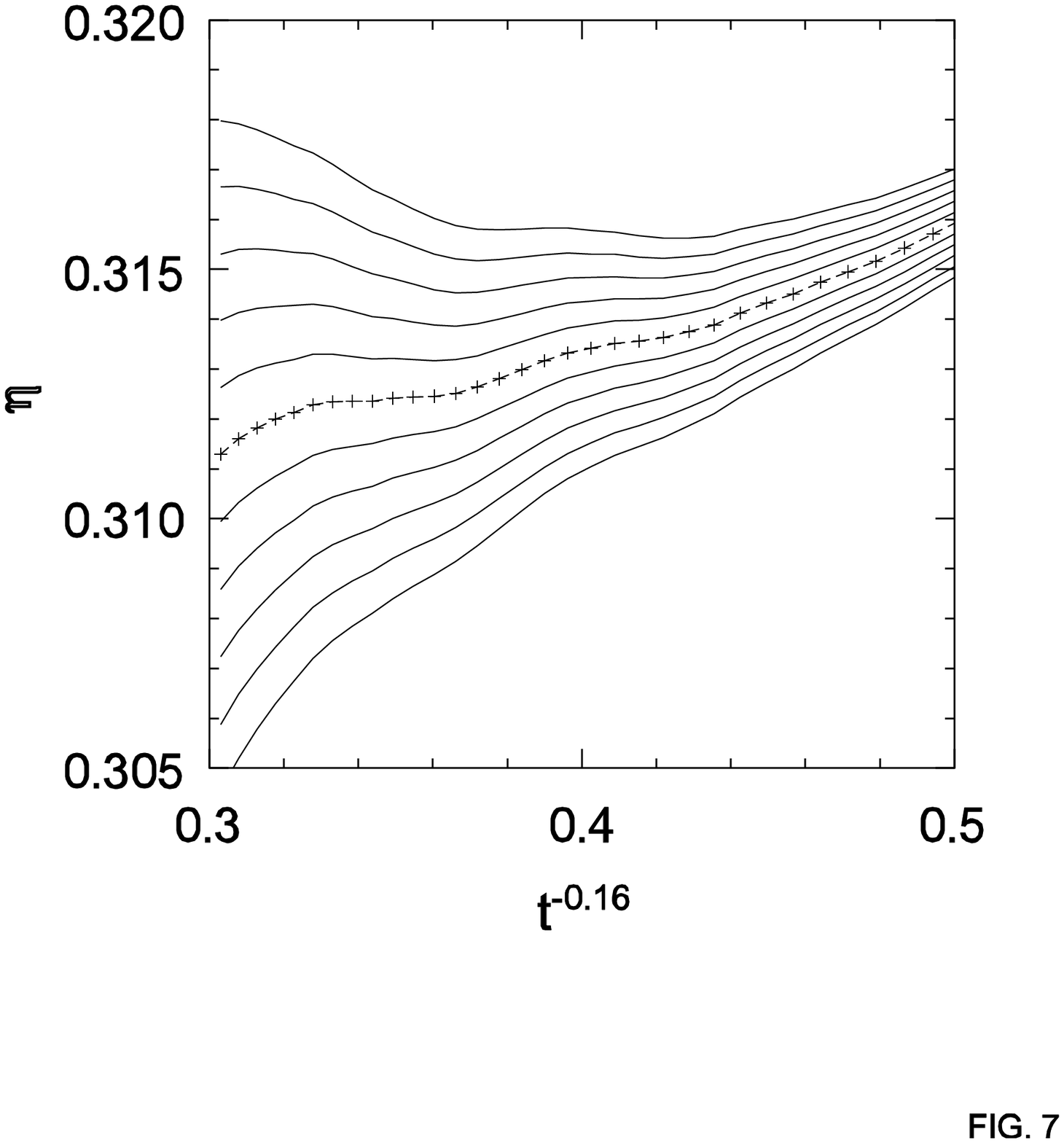,height=6.0in,width=4.5in}}
\end{figure}

\begin{figure} 
\centerline{\epsfig{file=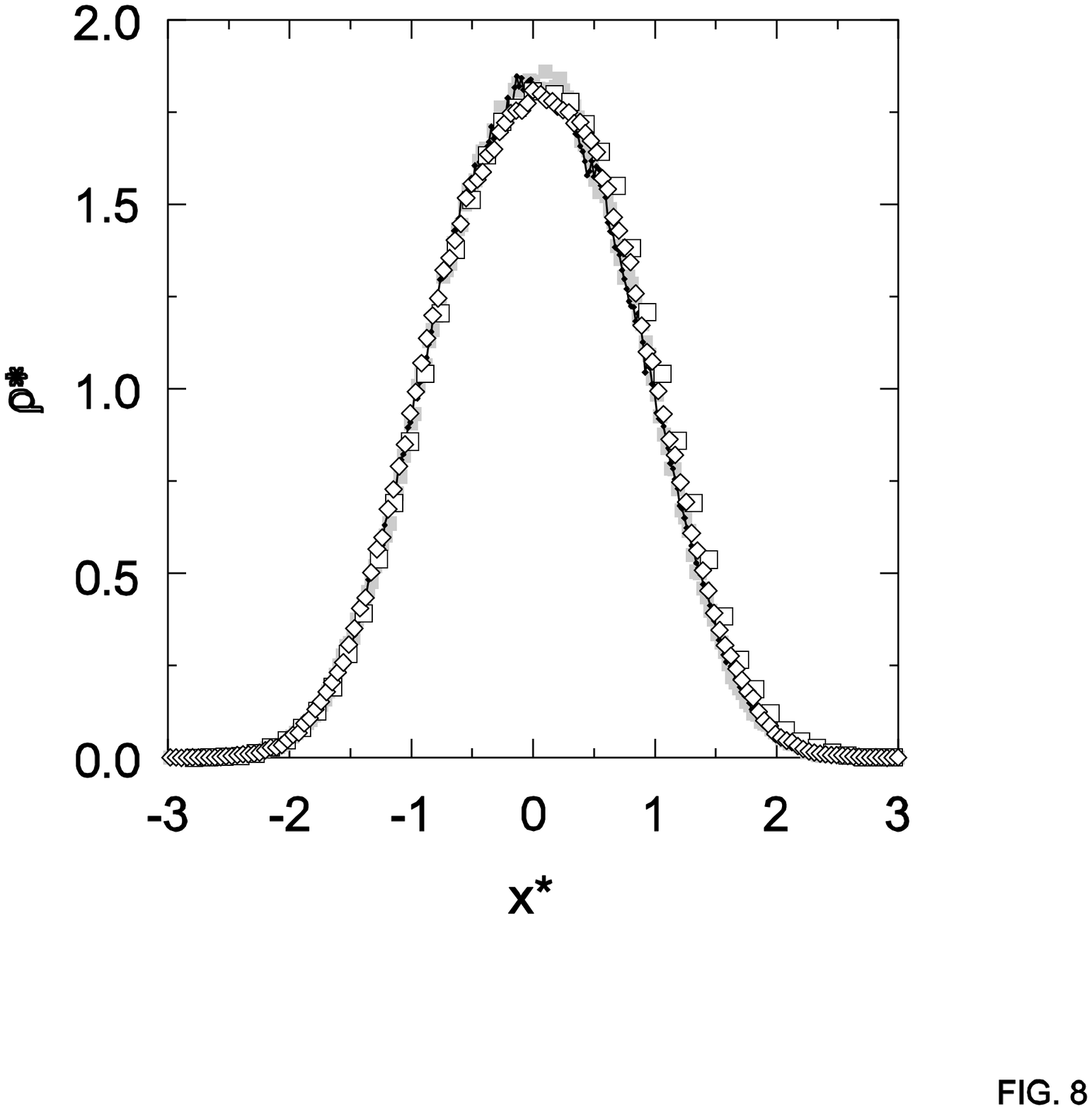,height=6.0in,width=4.5in}}
\end{figure}

\begin{figure} 
\centerline{\epsfig{file=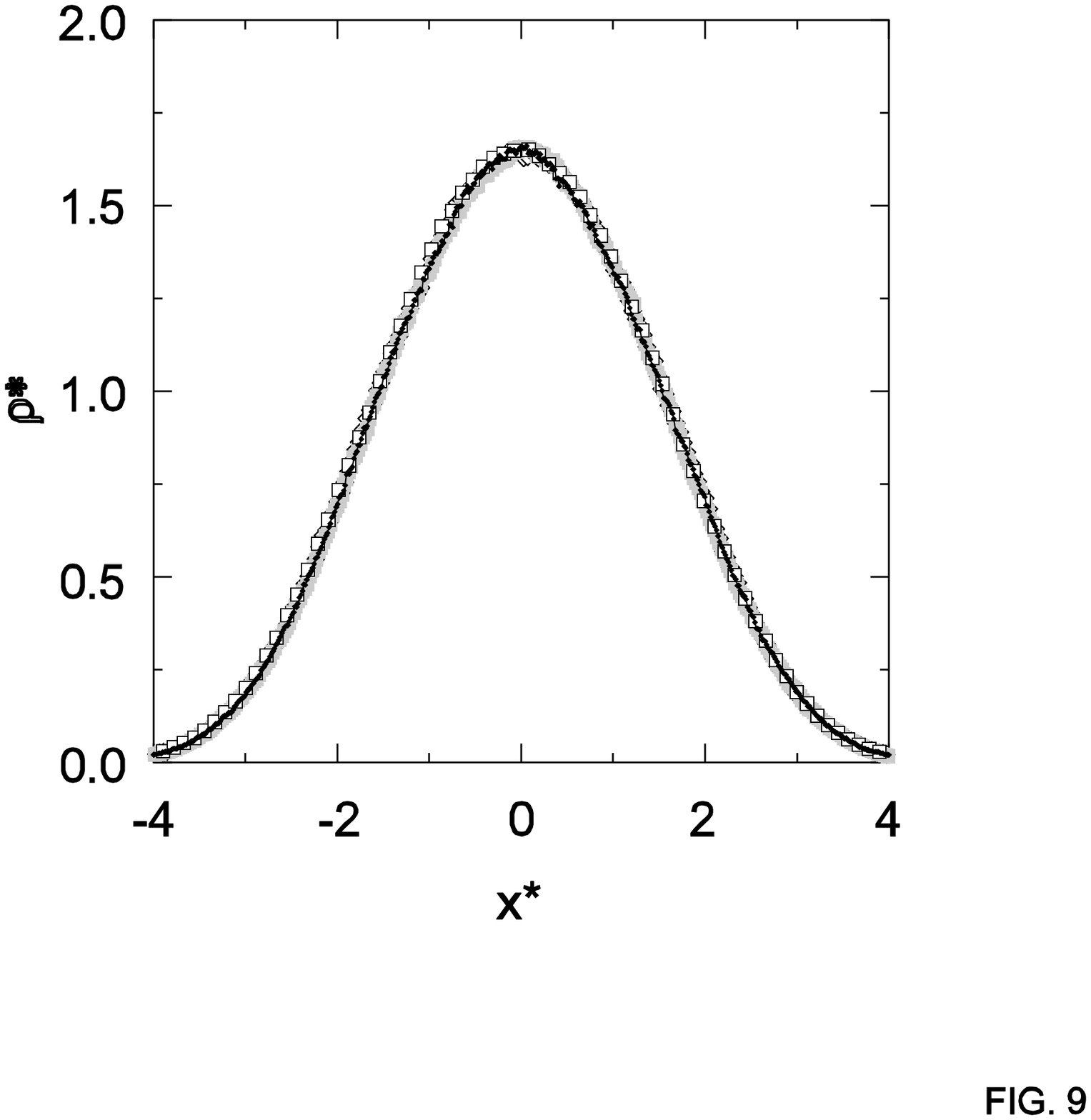,height=6.0in,width=4.5in}}
\end{figure}

\begin{figure} 
\centerline{\epsfig{file=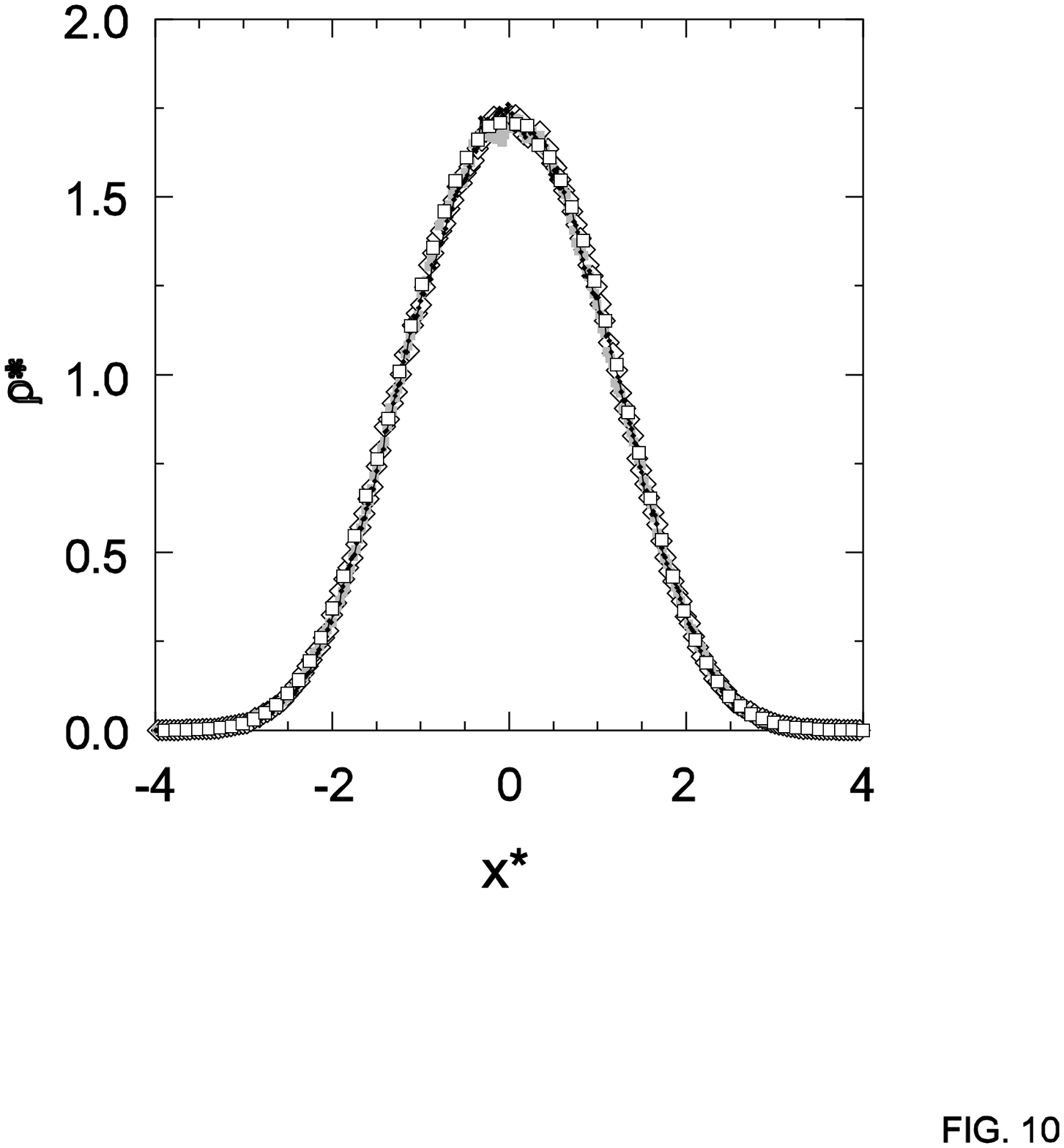,height=6.0in,width=4.5in}}
\end{figure}

\begin{figure} 
\centerline{\epsfig{file=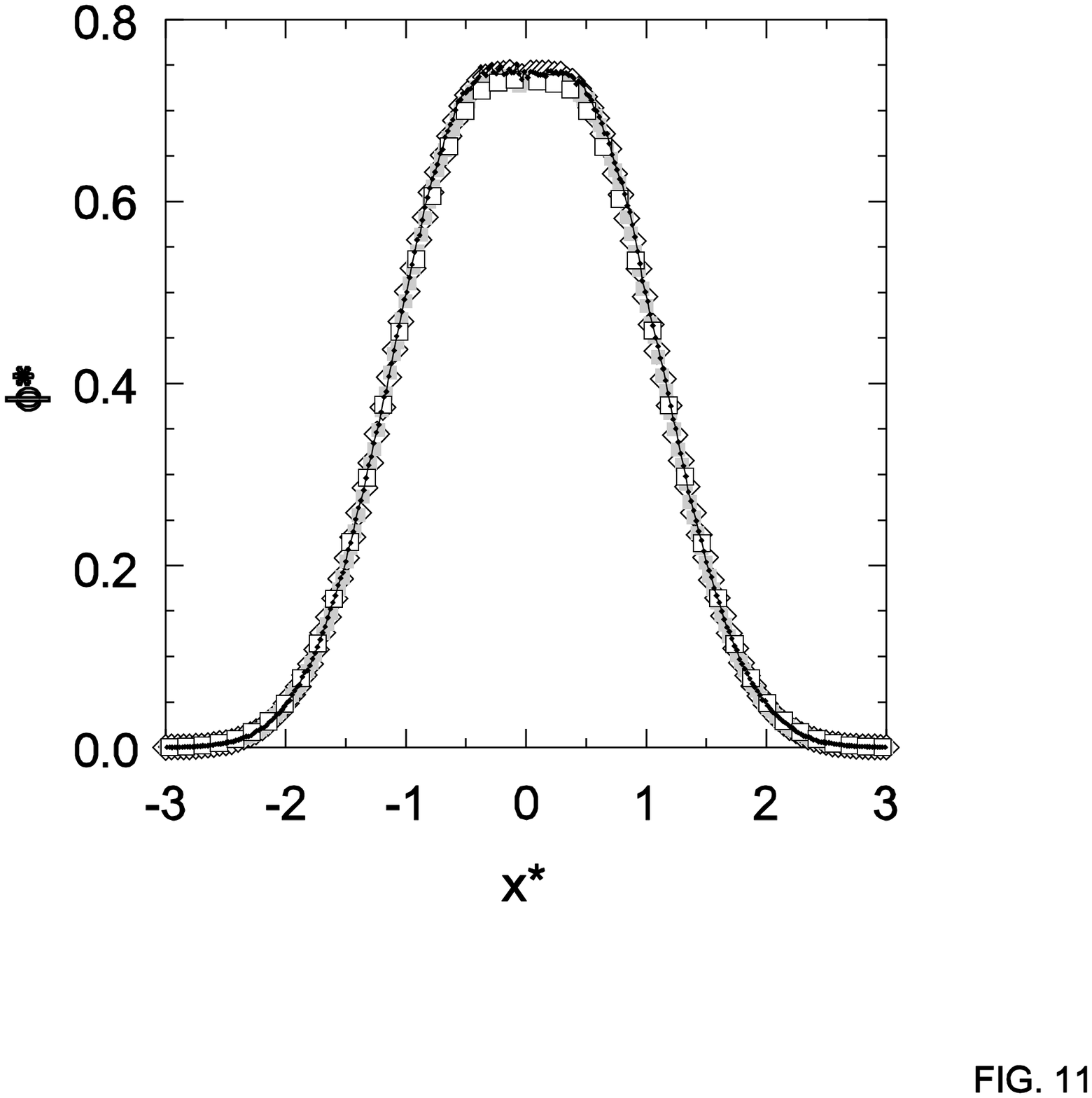,height=6.0in,width=4.5in}}
\end{figure}

\begin{figure} 
\centerline{\epsfig{file=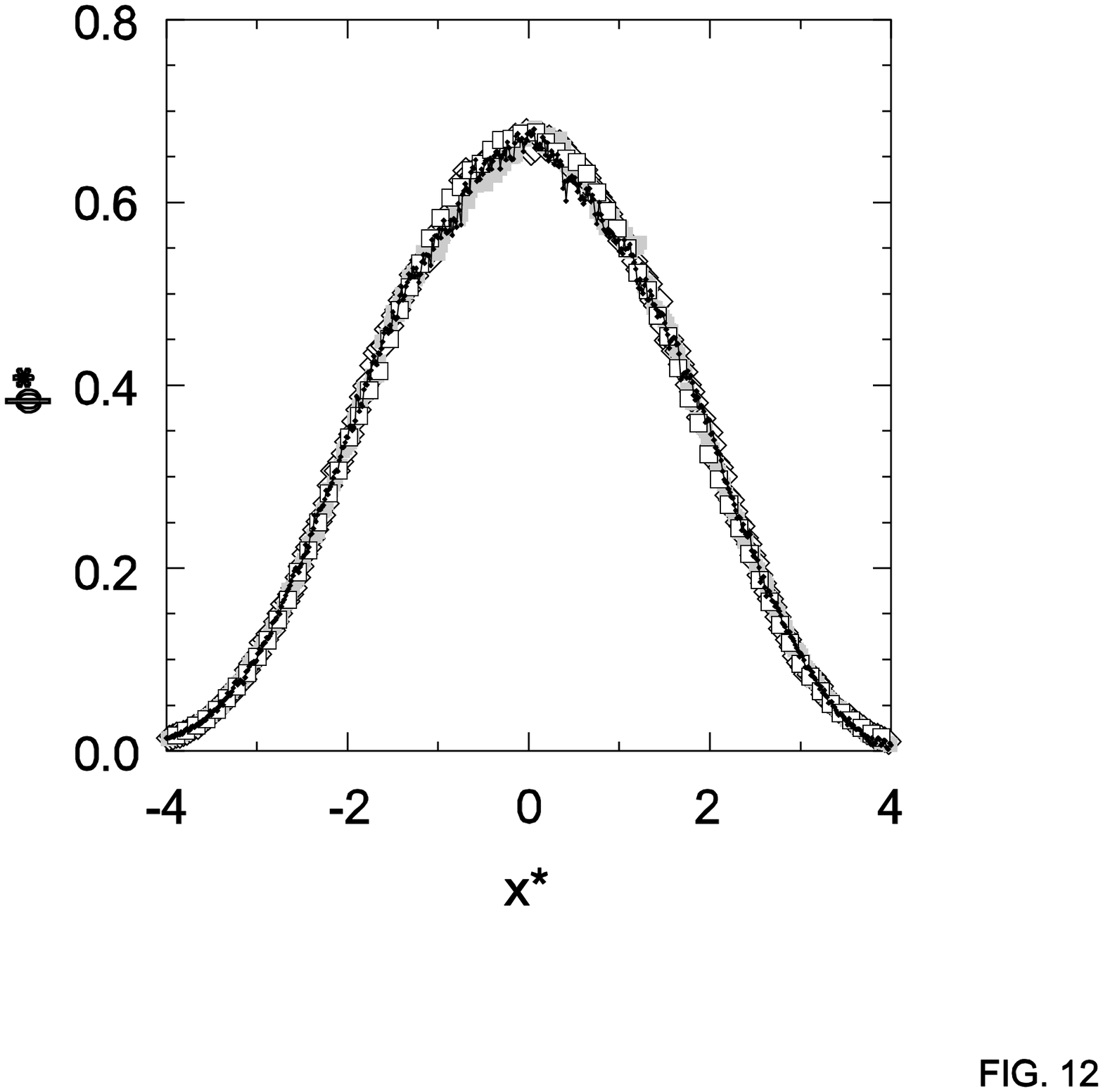,height=6.0in,width=4.5in}}
\end{figure}

\begin{figure} 
\centerline{\epsfig{file=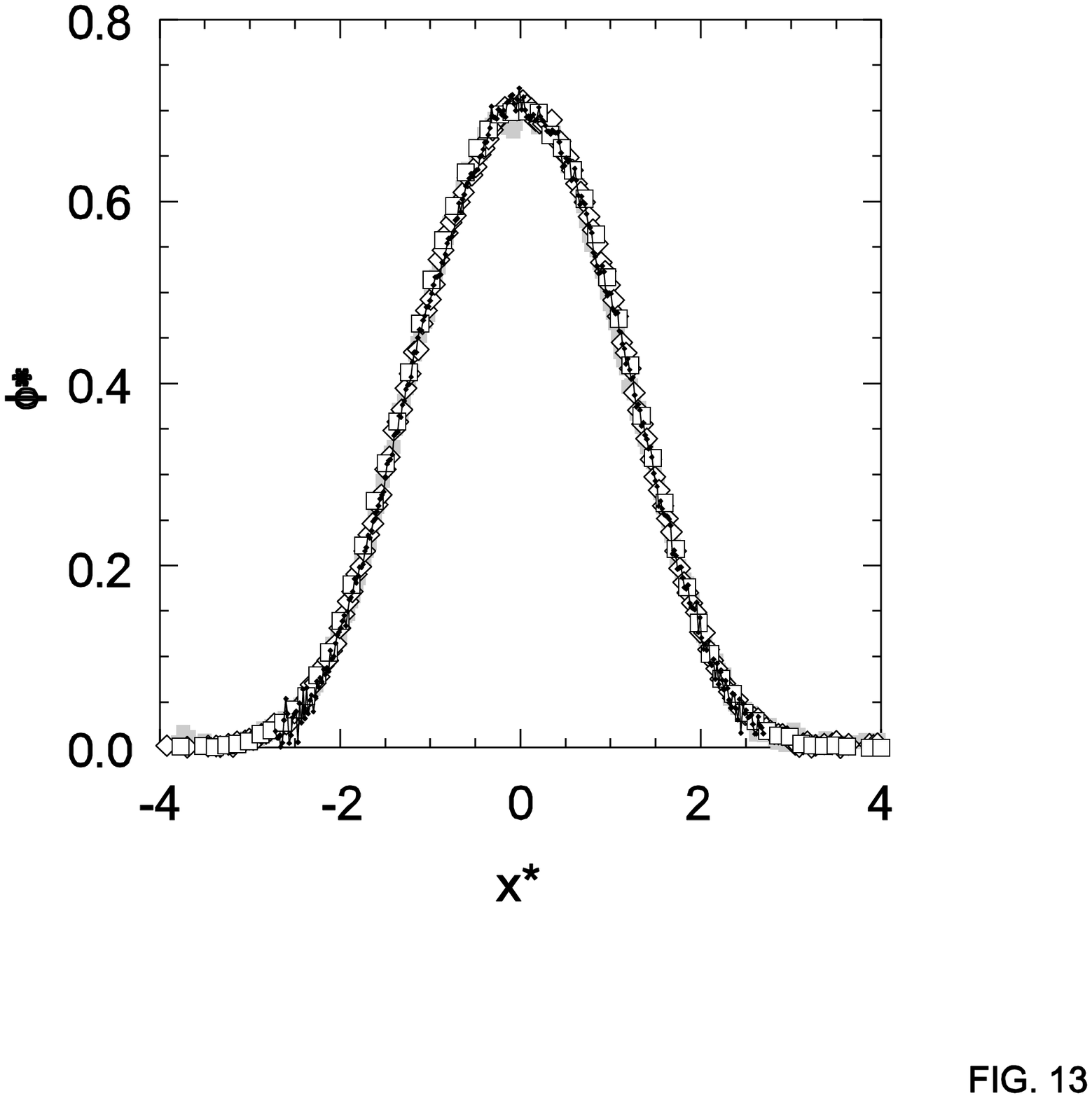,height=6.0in,width=4.5in}}
\end{figure}

\begin{figure} 
\centerline{\epsfig{file=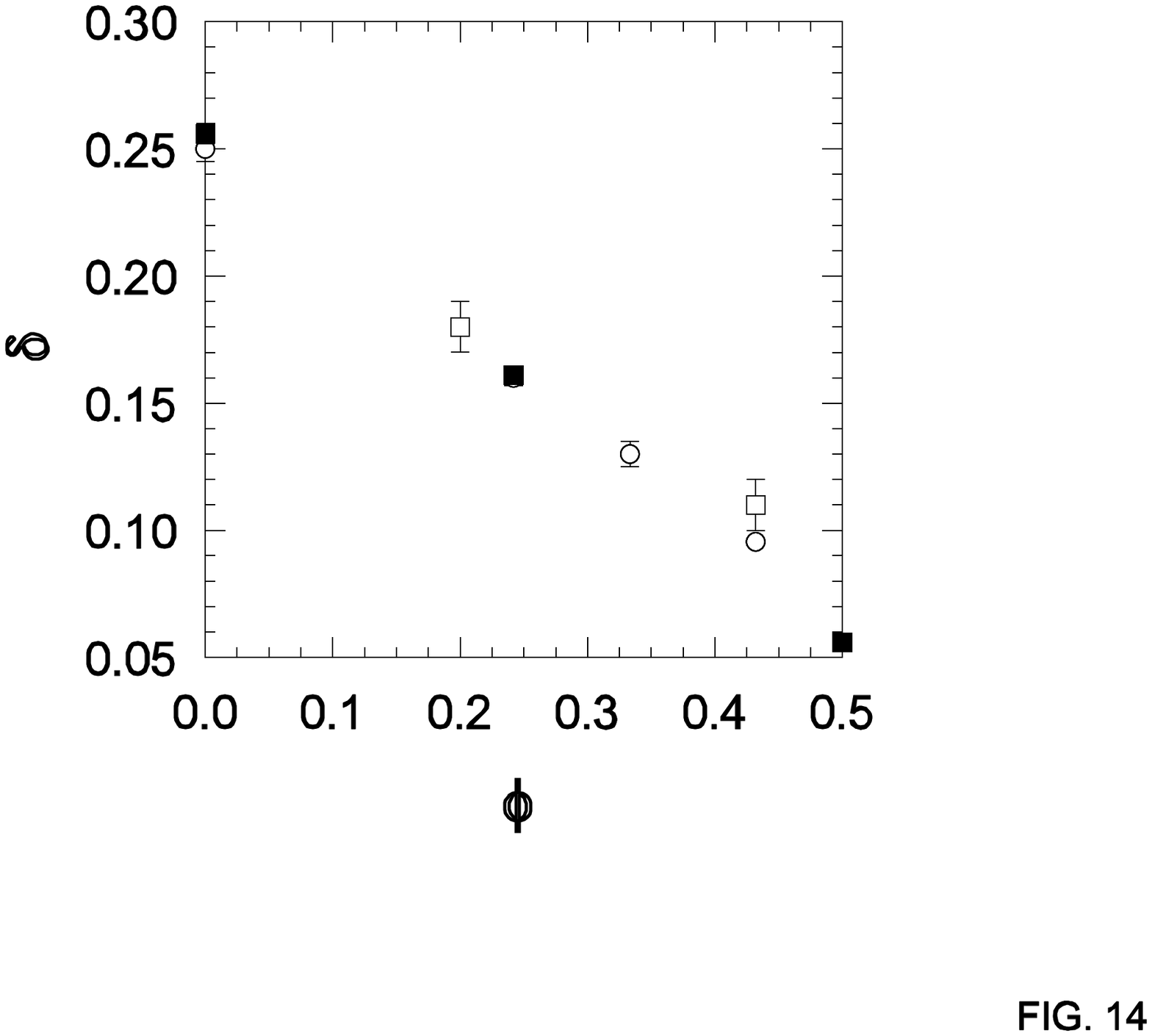,height=6.0in,width=4.5in}}
\end{figure}

\begin{figure} 
\centerline{\epsfig{file=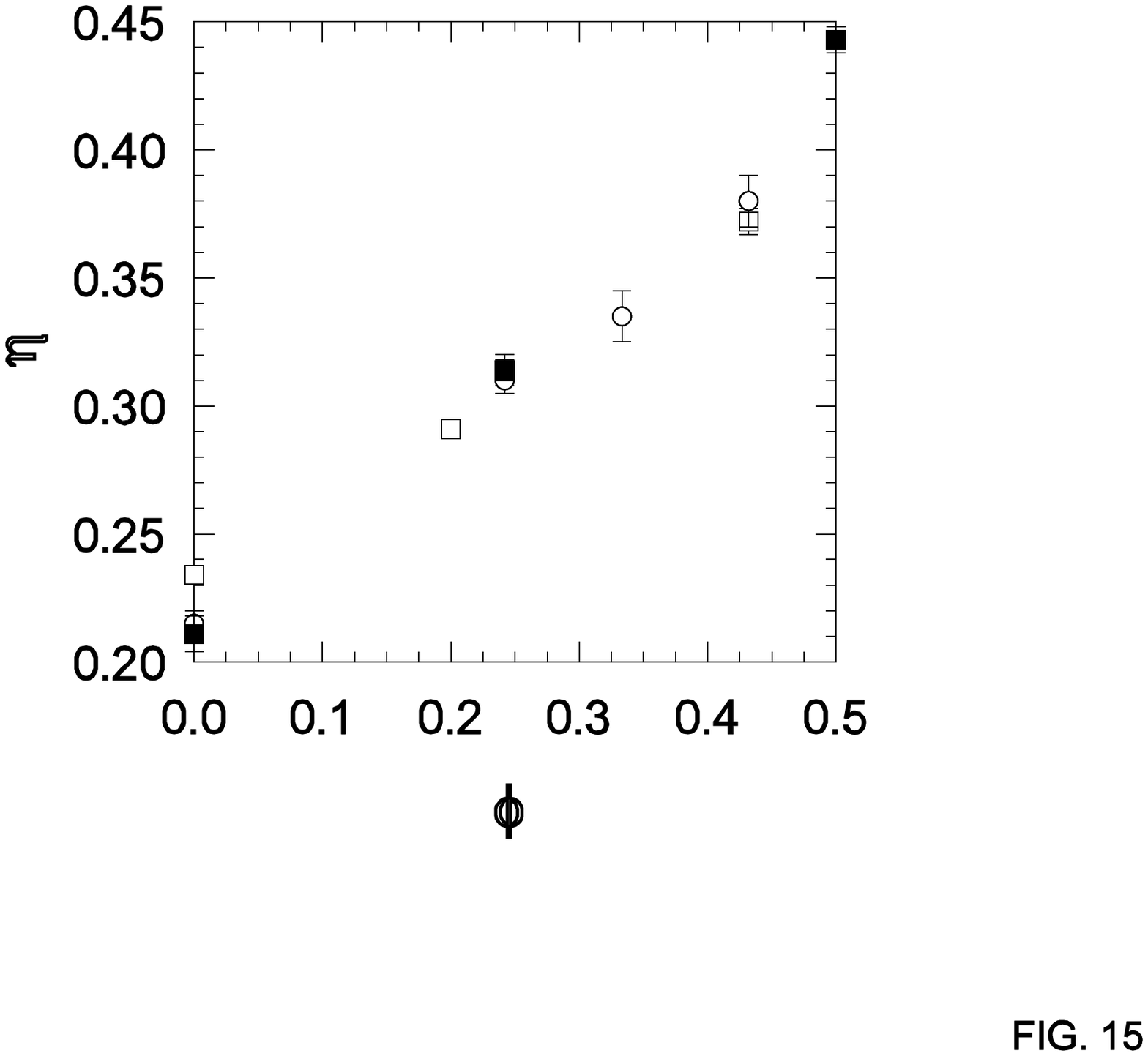,height=6.0in,width=4.5in}}
\end{figure}

\begin{figure} 
\centerline{\epsfig{file=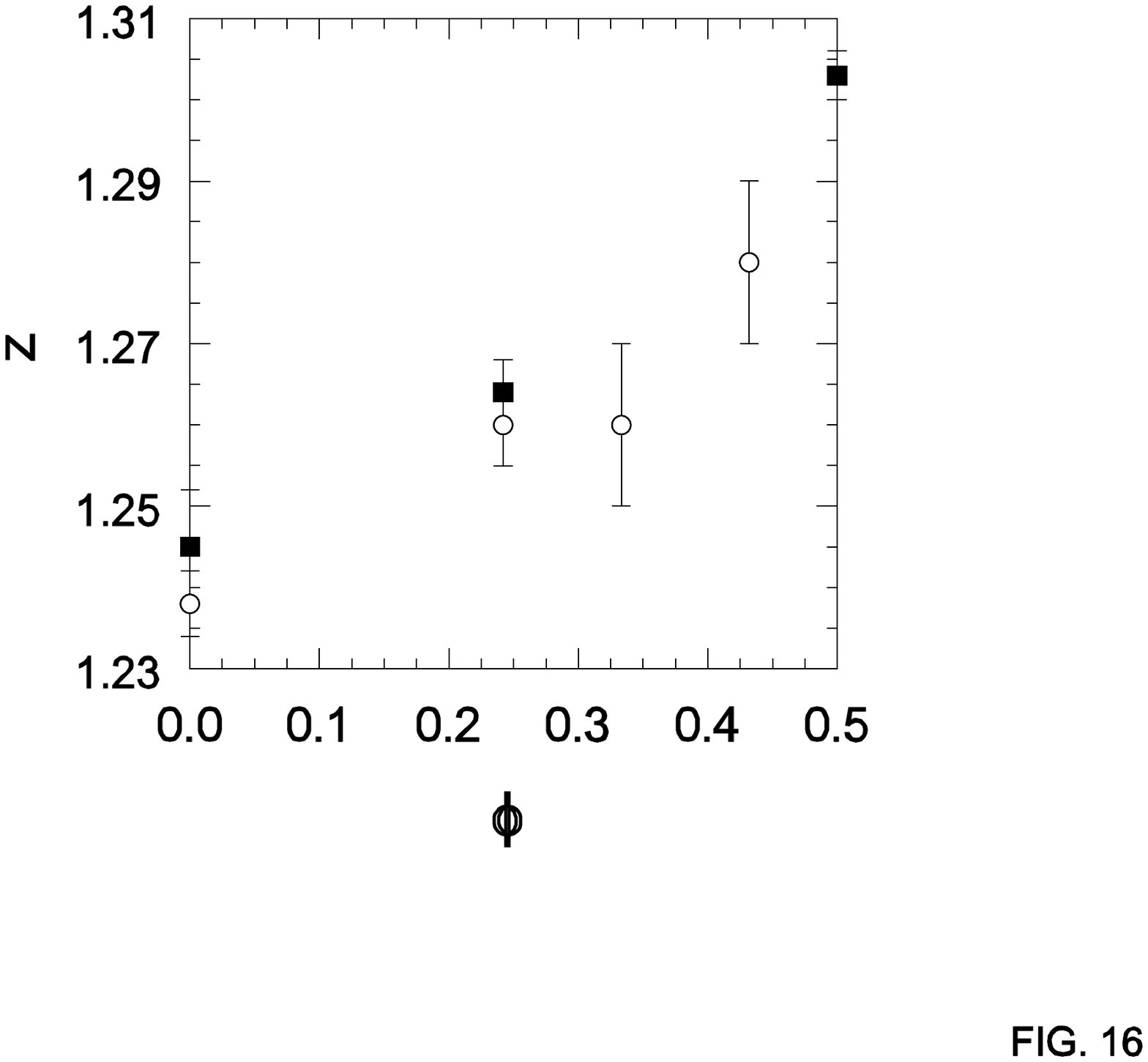,height=6.0in,width=4.5in}}
\end{figure}

\end{document}